# Time-Resolved Cathodoluminescence in an Ultrafast Transmission Electron Microscope

S. Meuret*[1], L. H.G. Tizei[2], F. Houdellier[1], S. Weber[1], Y. Auad[2], M. Tencé[2], H.-C. Chang[3], M. Kociak[2] and A. Arbouet*[1]

1 - CEMES-CNRS Université de Toulouse 29 rue Jeanne Marvig 31055 Toulouse France

2 - Université Paris-Saclay, CNRS, Laboratoire de Physique des Solides, 91405, Orsay, France

3 - Institute of Atomic and Molecular Sciences, Academia Sinica, Taipei 106, Taiwan

Sophie.meuret@cemes.fr , arnaud.arbouet@cemes.fr

*Ultra-fast transmission electron microscopy (UTEM) combines sub-picosecond time-resolution with the versatility of TEM spectroscopies. It allows one to study the dynamics of materials properties combining complementary techniques. However, until now, time-resolved cathodoluminescence, that is expected to give access to the optical properties dynamics, was still unavailable in a UTEM. In this paper we report time-resolved cathodoluminescence measurements in an ultrafast transmission electron microscope. We measured lifetime maps, with a 12 nm spatial resolution and sub-nanoseconds resolution, of nano-diamonds with a high density of NV center. This study paves the way to new applications of UTEM and to correlative studies of optically active nanostructures.*

Excited states in atomic, molecular or solid-state systems can decay through a great variety of non-radiative or radiative relaxation channels. Light emission, also called luminescence, associated with radiative pathways contain invaluable information on the properties of these systems and their dynamics. Measuring the lifetime of an excited state is therefore essential to understand complex relaxation pathways. The study of light emission has motivated over the course of history a large number of major instrumental developments. The developed techniques have long relied on optical excitation of the sample [1]–[7]. However, the limited spatial resolution of optical spectroscopies and the sub-wavelength length scale of several important relaxation processes stimulated the search for alternative strategies. The excellent spatial resolution of electron microscopes and the richness of the secondary signals produced by the interaction between fast electrons and matter motivated very early ambitious developments aimed at obtaining with the same instrument information on the chemical structure (X-ray), surface properties (secondary electrons) and optical properties ((near-)visible light). One of the advantages of Transmission Electron Microscopy (TEM) over Scanning Electron Microscopy (SEM) is the possibility to study the electron after its interaction with the material under study. This allows for instance to analyze the electron beam energy exchange with the material (electron energy loss spectroscopy : EELS), or to determine the crystallographic structure of the sample (diffraction imaging) as well as to quantitatively map the electric and magnetic fields (Electron holography). TEM is therefore one of the most versatile tools to study matter at the nanometer scale. In a TEM, information on the material optical properties can be gained through the study of the generated visible light via Cathodoluminescence (CL) or the analysis of the EELS in the low-loss region of the spectrum, the two techniques giving access to different optical quantities [8], [9].

The development of CL spectroscopy in a scanning TEM (STEM-CL) has been a major milestone [10]–[12] that allowed the study of solid state systems structured at the atomic scale, such as Light Emitting Diodes (LEDs), semiconductor lasers or color centers [13] at the relevant scale. STEM-CL was for example used to study the quantum Stark effect of III-V heterostructures [14], [15], to spatially resolve atomic defects in diamond [16], or reveal the connection between the formation of GaN/AlN quantum dots and threading dislocations [17]. It has become a well-established technique in nano-optics. However, the continuous electron beam of standard electron microscopes has long restricted cathodoluminescence to the study of steady state optical properties and prevented the measurement of the excited state lifetime or the carrier mobility. Recently, the measurement of the CL autocorrelation function allowed the measurement of the lifetime even with a continuous beam [18]–[20] but with a low signal to noise ratio (SNR) due to its dependence on the square of the intensity [21]. However, it demonstrated the possibility and interest of measuring lifetimes with sub 10-nm spatial resolution even in buried structures [12].

The search for time-resolved CL measurement using a pulsed electron excitation was therefore sought after early. Indeed, the first time-resolved cathodoluminescence study in an SEM was performed in the 80's. Christen and others [22]–[24], used a blanking scheme where the electron beam was swept over an aperture [25]–[27]. Despite a long pulse (> 100 ns) the sharp falling edge (< 1 ns) resulted in a sub-nanosecond resolution but at the expense of a limited spatial resolution (> 1 μm). It is only in 2005 that Merano et al [28] achieved the first time-resolved cathodoluminescence study combining picosecond temporal resolution and 50 nanometer spatial resolution. They used a laser-driven gold photocathode to generate a picosecond pulsed electron beam [29]–[31]. Since then, time-resolved cathodoluminescence in a SEM has been used to study the carrier relaxation dynamics [32] as well as the influence of basal stacking faults [33], strain [34] or exciton mobility [35] on the optical properties. Until now, time-resolved cathodoluminescence studies have only been performed in SEMs, limiting the achievable spatial resolution and material properties accessible from complementary signals.

In this letter we report the first time-resolved cathodoluminescence study performed in a pulsed electron gun TEM. We measured the spatial variation of the lifetime of atomic defects (nitrogen vacancy – NV – center) in nano-diamonds with a sub-nanosecond time resolution and 12 nm spatial resolution.

Figure 1-a,c shows a sketch of the experimental set-up ; an Hitachi HF2000 ultrafast transmission electron microscope (UTEM) equipped with an ultra-fast cold-field emission source described in previous articles [36]–[38]. A femtosecond laser generates a 400 fs pulsed electron beam from a sharp tungsten tip. The electrons are then accelerated to 150 keV. The repetition rate is set to 2 MHz. The pulsed electron beam can be focused on the sample within a probe of less than 1 nm, and scanned over the region of interest [37]. A parabolic mirror collects the cathodoluminescence generated by the interaction of the electron beam with the sample and sends it to a multimode optical fiber. The main challenge of cathodoluminescence in a UTEM is the signal to noise ratio as the current of the pulsed electron beam (about 100 of fA on the sample [39]) is much smaller than the one generated with a conventional TEM leading to a drastically decreased CL signal. The large collection angle of the parabolic mirror is therefore essential to collect efficiently the luminescence signal. It is followed by a lens whose numerical aperture is matched to an optical fiber redirecting the CL signal either to an optical spectrometer or to a Single Photon Counting Module (SPCM) connected to a correlator. The numerical matching ensures the lowest possible loss of photon flux, which is essential for these low signal experiments. Using the open-source acquisition software PyMoDAQ [40], we can record either the optical spectrum or the decay trace at each position of the electron beam. In this work, we studied nano-diamonds containing a large number of NV centers [41],

[42] with a mean diameter of about 100 nm. The NV center is an atomic defect in the diamond crystal lattice consisting of a substitutional nitrogen atom and a neighboring vacancy, it is a very well-known single photon emitter particularly stable under electron excitation and bright at room temperature[43], [44]. The NV center has two different charge states, the neutral $NV^0$ and the negatively charged $NV^{(-)}$. Cathodoluminescence spectroscopy is probing almost exclusively the $NV^0$ state [45]. Figure 1-b displays a typical cathodoluminescence spectrum of nano-diamonds used in this study. It was recorded with a continuous electron beam, at room temperature in a VG HB510 scanning TEM. The zero-phonon line emission is visible at 575 nm as well as the broad sideband associated with phonon-assisted emission from 575 to 800 nm.

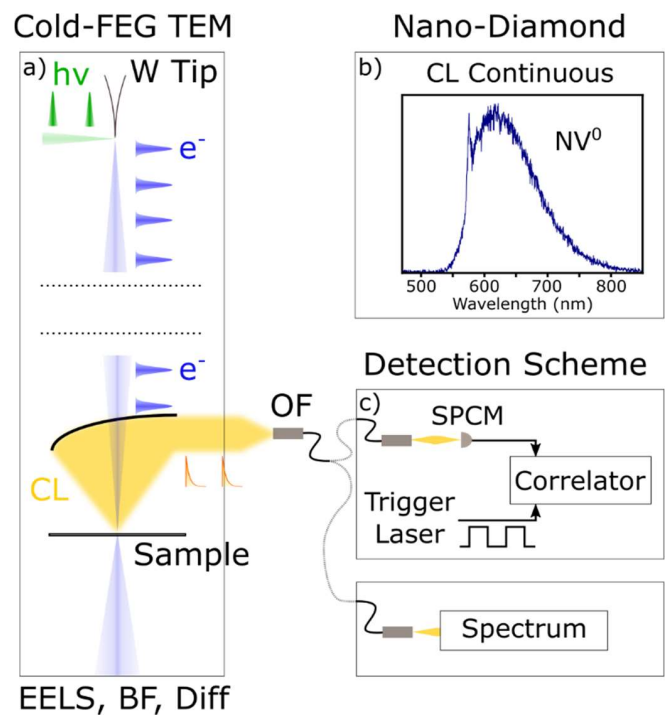

Figure 1: Schematics of the experimental set-up. a) In a modified cold-FEG Hitachi HF2000 microscope, a 2 MHz femto-second laser at 515 nm triggers the emission of femtosecond pulses of electrons from a Tungsten tip. A parabolic mirror collects the cathodoluminescence from the sample. c) Thanks to an optical fiber coupled to the parabolic mirror through a lens (not shown) the light goes either to a spectrometer or to a single photon counting module (SPCM). In the later, a correlator records a histogram of the delays between excitation (trigger) and CL photon detection. Such a histogram forms a decay trace from which the lifetime of the excited state can be extracted. b) Spectrum of a diamond $NV^0$ center, recorded in a VG510 scanning transmission electron. This spectrum was taken with a continuous beam at room temperature on the same batch of nano-diamonds used for this study. One can see the zero-phonon line at 575 nm followed by the phonon replica from 575 to 800 nm.

Figure 2 shows the spectrum and decay trace taken for a fixed position of the electron beam on a nano-diamond cluster. In order to maximize the SNR for the acquisition of the CL spectrum (Figure 2-a), no condenser aperture was inserted in the electron column to maximize the electron beam current. This induce a loss in spot size due to geometrical aberrations of condenser lenses. The spectrum acquisition time was 30 s. Nevertheless, we can notice that there is still enough spatial resolution to observe a change in intensity between the two different positions (about 400 nm apart) as noted in the insert of Figure 2-a. The peak at 515 nm visible in the spectrum is mostly due to scattering from the 515 nm laser exciting the

Tungsten tip in the electron gun. Direct excitation of the nano-diamonds by the scattered laser light can however be discarded as (i) the number of photons travelling from the tungsten tip to the sample is obviously very low due small apertures size along the electron optics (mainly source extractor and differential aperture), (ii) a significant contribution from such direct optical excitation is not consistent with the observed variations of the signal with the electron beam position and (iii) direct optical excitation of the NV centers would yield a significant contribution from the negatively charged NV$^-$ charge state and was not observed. Figure 2-b displays the decay trace measured at the same position as the blue spectrum of Figure 2-a. To acquire this decay trace, we record the delay between the excitation (trigger: electronic signal sent from the laser) and the CL photon detection. Due to the correlated acquisition between the excitation and the detection, the SNR improves greatly and we can add an aperture to improve the probe size down to about 5 nm. A long path filter at 550 nm is added in front of the SPCM detector in order to avoid the laser tail visible on the spectrum (Figure 2-a). In these conditions, we detect 40 photons (above the dark noise) per second on the APD. The decay trace displayed in Figure 2-b is taken with an integration time of 200 s.

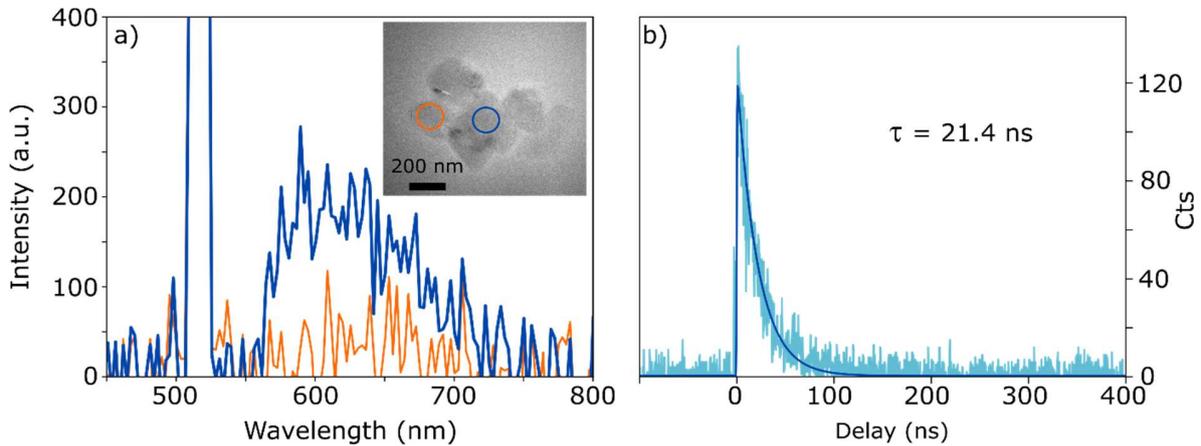

*Figure 2 : Spectrum and decay trace measured in the HF2000 based UTEM a) Average spectrum of a nano-diamond cluster (TEM image in Inset). The integration time was 30 s, and no illumination aperture was used to maximize the signal. b) Decay trace recorded for 200 s, the electron probe (5 nm spot using a 70 µm illumination aperture) excited the area highlighted by the blue circle in the inset of a). An exponential fit (thick blue line) gives a lifetime τ of 21.4 ns in the expected range for nano-diamond lifetimes.*

In order to extract the lifetime from the decay trace we fit the data with the convolution of an exponential decay and a gaussian. The gaussian function accounts for the time resolution of the experiment. It could also account for a slowly increasing contribution associated to the diffusion of carriers but in the case of nano-diamonds excited by 150 keV electrons we can safely consider the excitation of the defect centers instantaneous compared to the nanosecond lifetime. The instrument response function of the experiment is defined as IRF = $\sqrt{\sum \sigma_i^2}$ where $\sigma_i$ accounts for the different sources of error. In our case the contribution from the electron pulse-width (about 400 fs) is negligible, the time bin is set to 400 ps and the photon detector has a resolution of 350 ps. We therefore expect an IRF of about 530 ps. In the case of Figure 2-b, the fit yields an IRF of 873 ps which is close to what we expect considering the histogram resolution. The lifetime found for the decay trace displayed in Figure 2-b is 21.4 ± 0.5 ns, in the range of expected values for nano-diamonds [46].

These nano-diamonds contain much more than one center. The carrier diffusion length being of the order of 50 nm [16], we expect that the electron beam excites all the centers whatever its position in the nano-diamond. However, the excitation strength of a given center depends on (i) its position inside the diamond and (ii) its position relative to the electron beam. The measured lifetime is therefore an average of the lifetimes of all the centers excited by the electron beam. Nevertheless, lifetime correlation-based measurements performed on similar diamonds have shown a strong disparity of the averaged lifetimes measured between objects [47].

Therefore, we show in Figure 3 the results of spatially-resolved lifetime measurements performed with Time resolved STEM-CL. We scanned the electron beam over the nano-diamonds cluster and recorded a decay trace at each position. Using the same STEM aperture as for the decay trace shown in Figure 2-b (5 nm probe), we scanned a 200 nm region with square pixels of 6 nm width and an integration time of 5s per pixel. Figure 3 – b shows the intensity map extracted from the decay trace map by summing, for each pixel, all the histogram bin of a 500 ns window (2 MHz repetition rate). We can see quite clearly the intensity variations over the scan depending on the electron beam position. Before fitting each pixel content with the same function as for the average lifetime of Figure 2-b we binned four adjacent pixels to increase the signal to noise ratio. The equivalent integration time is therefore 20 s for each 12 nm width pixel. The lifetime extracted from the fit at each pixel is shown in Figure 3-c. The missing pixels represent pixels lying outside of the diamonds where the signal was too low to perform an accurate fit. Figure 3-d shows two decay traces acquired at different locations on the nano-diamonds and their corresponding fit. The measured lifetime depends strongly on the electron beam position and decreases from 23.4 ns to 15.8 ns when the electron beam is displaced by less than 50 nm. The signal to noise ratio (SNR) of these spatially-resolved data is worse than in the case of Figure 2. We measured lifetime errors between 0.5 and 1 ns for most pixels and up to 2 ns for the noisiest pixel (See SI). Other examples of lifetime maps on nano-diamonds are given in the supplementary together with more details about the fitting procedure.

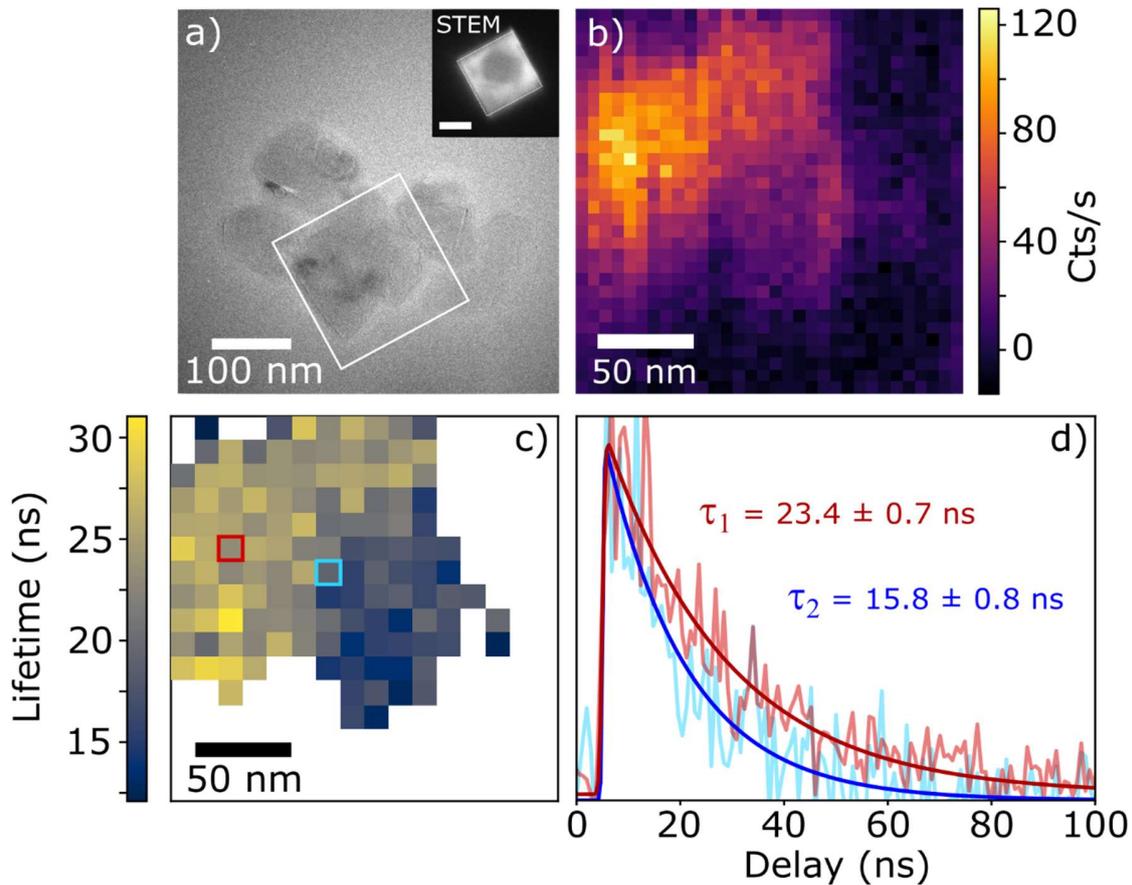

*Figure 3 : Lifetime mapping on a Nano-diamond cluster. a) TEM images of the cluster, the white square represents the scanned area. Inset: Scanning probe over the region of interest taken with the TEM CCD camera. scale bar 100 nm b) Intensity map extracted from the decay trace map by summing all the histogram bins of a 500 ns window, pixel size 6.3 nm. c) Lifetime map after binning four adjacent pixels (effective pixel size 12.4 nm). At each pixel a fit of the decay trace has been done and the extracted lifetime reported on the map. Pixels where the signal was too low to extract a lifetime are masked. D) Decay traces and corresponding fit at the locations marked by the blue and red squares in c).*

To conclude, we have performed the first time-resolved cathodoluminescence experiments in an Ultrafast Transmission Electron Microscope. Due to its high brightness cold-field emission source, this instrument can focus femtosecond electron pulses in a nanometric probe maintaining an optimum brightness. The light emitted from individual emitters is efficiently collected using a high numerical aperture parabolic mirror located close to the sample. This allowed to map the lifetime of NV centers in nano-diamonds with a spatial resolution of about 12 nm and a sub-nanosecond temporal resolution. Such a spatially-resolved lifetime measurements performed using an UTEM are extremely promising to investigate light emission dynamics from complex nanoscale systems [20]. In particular, the possibility of combining information on the emission dynamics together with structural, morphological, or chemical analysis at the atomic scale will bring invaluable information on the physics of the atomic scale light sources that play an increasing role in opto-electronic devices. This study opens a new application domain for Ultrafast Transmission Electron Microscopes and a wealth of possible experiments combining optical and electron excitation of nanoscale systems.

Acknowledgment:

This project has been funded in part by the European Union's Horizon 2020 research and innovation program under grant agreement No 823717 (ESTEEM3) and 101017720 (EBEAM). This project has been funded in part by the ANR under the grant agreement ANR-19-CE30-0008 ECHOMELO and grant agreement ANR-14-CE26-0013 FemtoTEM.

Conflict of interest:

MK patented and licensed the technologies used in this work for light collection to Attolight for which he is a part time consultant at Attolight. All other authors declare no competing interests.

**References:**

[1] K.-D. Park *et al.*, "Time-Resolved Ultraviolet Near-Field Scanning Optical Microscope for Characterizing Photoluminescence Lifetime of Light-Emitting Devices," *J. Nanosci. Nanotechnol.*, vol. 13, no. 3, 2013.

[2] A. M. Mintairov *et al.*, "High-spatial-resolution near-field photoluminescence and imaging of whispering-gallery modes in semiconductor microdisks with embedded quantum dots," *Phys. Rev. B - Condens. Matter Mater. Phys.*, vol. 77, no. 19, pp. 1–7, 2008.

[3] E. Gaviola, "Ein Fluorometer. Apparat zur Messung von Fluoreszenzabklingungszeiten," *Zeitschrift für Phys.*, vol. 42, no. 11–12, pp. 853–861, 1927.

[4] B. D. Venetta, "Microscope Phase Fluorometer for Determining the Fluorescence Lifetimes of Fluorochromes," *Rev. Sci. Instrum.*, vol. 30, no. 6, pp. 450–457, 1959.

[5] J. R. Lakowicz, "Frequency-Domain Lifetime Measurements," in *Principles of Fluorescence Spectroscopy*, J. R. Lakowicz, Ed. Boston, MA: Springer US, 2006, pp. 157–204.

[6] I. Bugiel, K. Konig, and H. Wabnitz, "Investigation of cells by fluorescence laser scanning microscopy with subnanosecond time resolution," *Lasers in the Life Sciences*, vol. 3, no. 1. pp. 47–53, 1989.

[7] A. Gruber, A. Dräbenstedt, C. Tietz, L. Fleury, J. Wrachtrup, and C. Von Borczyskowski, "Scanning confocal optical microscopy and magnetic resonance on single defect centers," *Science (80-. ).*, vol. 276, no. 5321, pp. 2012–2014, 1997.

[8] B. G. Yacobi and D. B. Holt, *Cathodoluminescence Microscopy of Inorganic Solids*. New York: Springer, 1990.

[9] A. Losquin and M. Kociak, "Link between Cathodoluminescence and Electron Energy Loss Spectroscopy and the Radiative and Full Electromagnetic Local Density of States," *ACS Photonics*, vol. 2, no. 11, pp. 1619–1627, 2015.

[10] P. M. Petroff and D. V Lang, "A new spectroscopic technique for imaging the spatial distribution of nonradiative defects in a scanning transmission electron microscope," *Appl. Phys. Lett.*, vol. 31, no. 2, pp. 60–62, 1977.

[11] N. Yamamoto, K. Araya, and F. García de Abajo, "Photon emission from silver particles induced by a high-energy electron beam," *Phys. Rev. B*, vol. 64, no. 20, p. 205419, Nov. 2001.

[12] M. Kociak and L. F. Zagonel, "Cathodoluminescence in the scanning transmission electron microscope," *Ultramicroscopy*, vol. 176, pp. 112–131, 2017.

[13] T. Coenen and N. M. Haegel, "Cathodoluminescence for the 21st century: Learning more from light," *Appl. Phys. Rev.*, vol. 4, no. 3, p. 031103, 2017.

[14] L. F. Zagonel et al., "Nanometer scale spectral imaging of quantum emitters in nanowires and its correlation to their atomically resolved structure.," *Nano Lett.*, vol. 11, no. 2, pp. 568–73, Feb. 2011.

[15] J. T. Griffiths et al., "Nanocathodoluminescence Reveals Mitigation of the Stark Shift in InGaN Quantum Wells by Si Doping," *Nano Lett.*, vol. 15, no. 11, pp. 7639–7643, 2015.

[16] L. H. G. Tizei and M. Kociak, "Spectrally and spatially resolved cathodoluminescence of nanodiamonds: local variations of the NV(0) emission properties.," *Nanotechnology*, vol. 23, no. 17, p. 175702, 2012.

[17] G. Schmidt et al., "Direct evidence of single quantum dot emission from GaN islands formed at threading dislocations using nanoscale cathodoluminescence: A source of single photons in the ultraviolet," *Appl. Phys. Lett.*, vol. 106, no. 25, 2015.

[18] S. Meuret et al., "Lifetime Measurements Well below the Optical Diffraction Limit," *ACS Photonics*, vol. 3, no. 7, pp. 1157–1163, 2016.

[19] S. Finot et al., "Carrier dynamics near a crack in GaN microwires with AlGaN multiple quantum wells," *Appl. Phys. Lett.*, vol. 117, no. 22, p. 221105, 2020.

[20] S. Yanagimoto, N. Yamamoto, T. Sannomiya, and K. Akiba, "Purcell effect of nitrogen-vacancy centers in nanodiamond coupled to propagating and localized surface plasmons revealed by photon-correlation cathodoluminescence," *Phys. Rev. B*, vol. 103, no. 20, p. 205418, 2021.

[21] S. Meuret et al., "Complementary cathodoluminescence lifetime imaging configurations in a scanning electron microscope," *Ultramicroscopy*, vol. 197, no. November 2018, pp. 28–38, 2019.

[22] J. Christen and D. Bimberg, "Cathodoluminescence Imaging of Semiconductor Interfaces," *Oyo Buturi*, vol. 57, pp. 69–77, 1988.

[23] H. T. Lin, D. H. Rich, A. Konkar, P. Chen, and A. Madhukar, "Carrier relaxation and recombination in GaAs/AlGaAs quantum heterostructures and nanostructures probed with time-resolved cathodoluminescence," *J. Appl. Phys.*, vol. 81, no. 7, pp. 3186–3195, 1997.

[24] A. Bell et al., "Localization versus field effects in single InGaN quantum wells Localization versus field effects in single InGaN quantum wells," *Appl. Phys. Lett.*, vol. 84, no. 58, pp. 1–4, 2004.

[25] D. Winkler, R. Schmitt, M. Brunner, and B. Lischke, "Flexible picosecond probing of integrated circuits with chopped electron beams," *IBM J. Res. Dev.*, vol. 34, no. 2.3, pp. 189–203, 1990.

[26] K. URA, H. FUJIOKA, and T. HOSOKAWA, "Picosecond Pulse Stroboscopic Scanning Electron Microscope," *J. Electron Microsc. (Tokyo).*, vol. 27, no. 4, pp. 247–252, 1978.

[27] R. J. Moerland, I. G. C. Weppelman, M. W. H. Garming, P. Kruit, and J. P. Hoogenboom, "Time-

resolved cathodoluminescence microscopy with sub-nanosecond beam blanking for direct evaluation of the local density of states," *Opt. Express*, vol. 24, no. 21, p. 24760, 2016.

[28] M. Merano *et al.*, "Probing carrier dynamics in nanostructures by picosecond cathodoluminescence.," *Nature*, vol. 438, no. 7067, pp. 479–82, Nov. 2005.

[29] H. E. Elsayed-Ali and J. W. Herman, "Ultrahigh vacuum picosecond laser-driven electron diffraction system," *Rev. Sci. Instrum.*, vol. 61, no. 6, pp. 1636–1647, 1990.

[30] P. May, J. -M. Halbout, and G. Chiu, "Picosecond photoelectron scanning electron microscope for noncontact testing of integrated circuits," *Appl. Phys. Lett.*, vol. 51, no. 2, pp. 145–147, 1987.

[31] D. Yang, O. F. Mohammed, and A. H. Zewail, "Scanning ultrafast electron microscopy," vol. 2010, 2010.

[32] W. Liu *et al.*, "Spatially dependent carrier dynamics in single InGaN/GaN core-shell microrod by time-resolved cathodoluminescence," *Appl. Phys. Lett.*, vol. 112, no. 5, p. 52106, 2018.

[33] P. Corfdir *et al.*, "Exciton recombination dynamics in a -plane (Al,Ga)N/GaN quantum wells probed by picosecond photo and cathodoluminescence," *J. Appl. Phys.*, vol. 107, p. 43524, 2010.

[34] X. Fu *et al.*, "Exciton Drift in Semiconductors under Uniform Strain Gradients: Application to Bent ZnO Microwires," *ACS Nano*, vol. 8, no. 4, pp. 3412–3420, 2014.

[35] M. Shahmohammadi, G. Jacopin, X. Fu, J.-D. Ganière, D. Yu, and B. Deveaud, "Exciton hopping probed by picosecond time-resolved cathodoluminescence," *Appl. Phys. Lett.*, vol. 107, no. 14, p. 141101, 2015.

[36] F. Houdellier, G. M. Caruso, S. Weber, M. Kociak, and A. Arbouet, "Development of a high brightness ultrafast Transmission Electron Microscope based on a laser-driven cold field emission source," *Ultramicroscopy*, vol. 186, pp. 128–138, 2018.

[37] G. M. Caruso, F. Houdellier, S. Weber, M. Kociak, and A. Arbouet, "High brightness ultrafast transmission electron microscope based on a laser-driven cold-field emission source: principle and applications," *Adv. Phys. X*, vol. 4, no. 1, p. 1660214, 2019.

[38] G. M. Caruso, F. Houdellier, P. Abeilhou, and A. Arbouet, "Development of an ultrafast electron source based on a cold-field emission gun for ultrafast coherent TEM," *Appl. Phys. Lett.*, vol. 111, no. 2, 2017.

[39] F. Houdellier, G. M. Caruso, S. Weber, M. J. Hÿtch, C. Gatel, and A. Arbouet, "Optimization of off-axis electron holography performed with femtosecond electron pulses," *Ultramicroscopy*, vol. 202, pp. 26–32, 2019.

[40] S. J. Weber, "PyMoDAQ: An open-source Python-based software for modular data acquisition," *Rev. Sci. Instrum.*, vol. 92, no. 4, p. 45104, 2021.

[41] L.-J. Su *et al.*, "Creation of high density ensembles of nitrogen-vacancy centers in nitrogen-rich type Ib nanodiamonds.," *Nanotechnology*, vol. 24, no. 31, p. 315702, Aug. 2013.

[42] J. Botsoa *et al.*, "Optimal conditions for NV- center formation in type-1b diamond studied using photoluminescence and positron annihilation spectroscopies," *Phys. Rev. B*, vol. 84, no. 12, p. 125209, Sep. 2011.

[43]  R. Schirhagl, K. Chang, M. Loretz, and C. L. Degen, "Nitrogen-Vacancy Centers in Diamond: Nanoscale Sensors for Physics and Biology," *Annu. Rev. Phys. Chem.*, vol. 65, no. 1, pp. 83–105, 2014.

[44]  A. Mohtashami and a Femius Koenderink, "Suitability of nanodiamond nitrogen–vacancy centers for spontaneous emission control experiments," *New J. Phys.*, vol. 15, no. 4, p. 043017, Apr. 2013.

[45]  M. Solà-Garcia, S. Meuret, T. Coenen, and A. Polman, "Electron-induced state conversion in diamond NV centers measured with pump-probe cathodoluminescence spectroscopy," vol. 0, pp. 20–22, Oct. 2019.

[46]  H. Lourenço-Martins *et al.*, "Probing Plasmon-NV0 Coupling at the Nanometer Scale with Photons and Fast Electrons," *ACS Photonics*, vol. 5, no. 2, pp. 324–328, 2018.

[47]  H. Lourenço-Martins *et al.*, "Probing Plasmon-NV 0 Coupling at the Nanometer Scale with Photons and Fast Electrons," *ACS Photonics*, 2017.

# Supplementary Information: Time-Resolved Cathodoluminescence in an Ultrafast Transmission Electron Microscope

S. Meuret*[1], L. H.G. Tizei[2], F. Houdellier[1], S. Weber[1], Y. Auad[2], M. Tencé[2], H.-C. Chang[3], M. Kociak[2] and A. Arbouet*[1]

1 - CEMES-CNRS Université de Toulouse 29 rue Marvig 31055 Toulouse France

2 - Université Paris-Saclay, CNRS, Laboratoire de Physique des Solides, 91405, Orsay, France

3 - Institute of Atomic and Molecular Sciences, Academia Sinica, Taipei 106, Taiwan

Sophie.meuret@cemes.fr, arnaud.arbouet@cemes.fr

1. Lifetime Map Error and Fit precision

The fitting procedure is based on the convolution of an exponential and a Gaussian :

$$A * \left( e^{-\frac{(t-t_0)}{\tau}} \times \frac{1}{\sigma\sqrt{2\pi}} e^{-\frac{(t-t_0)^2}{2\sigma^2}} \right) + B$$

With A the amplitude of the decay trace and B the background. T represents the lifetime and $t_0$ the time of the excitation. The full width at half maximum of the Gaussian is $2\sqrt{\ln(2)}\sigma \approx 2.35\sigma$ and represents the instrumental response function. The lifetime error was calculated with the square root of the covariance from the fit. Figure S1 shows the extracted lifetimes and associated errors for each pixels.

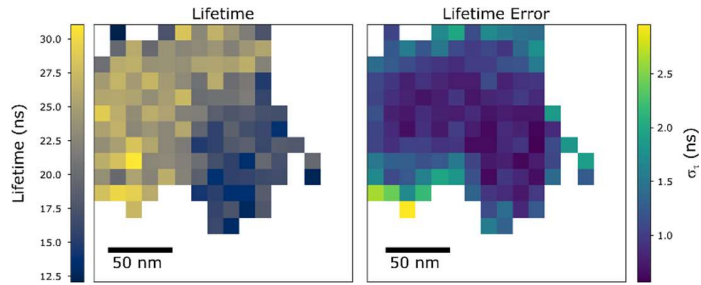

*Figure S4: Lifetime map of figure 3 of the main text with the associated error on the fit for the lifetime parameter calculated from the square root of the covariant.*

We checked by binning the data by 4 that the shorter lifetime where not simply due to a smaller SNR, as we can see on Figure S2 the lifetime stays consistent with the value found for the binning by 2.

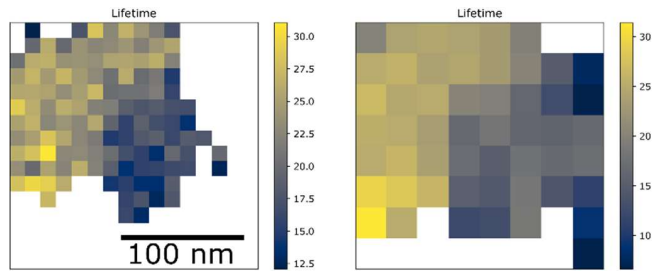

*Figure S5: Lifetime maps with a binning 2 (left) and 4 (right). The lifetime found is similar for both maps.*

2. Other Example of lifetime map on nanodiamond

The same experimental conditions were used for these two other nanodiamonds.

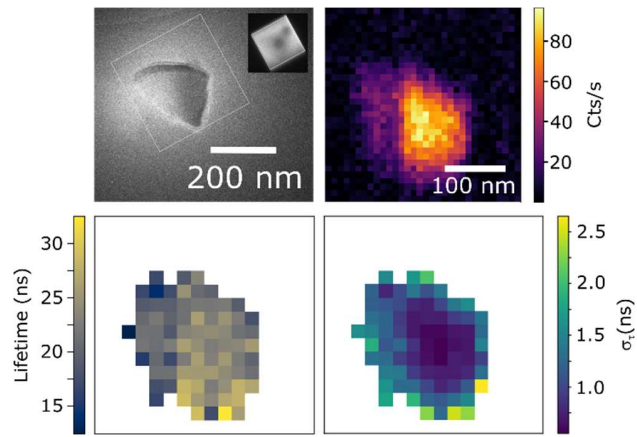

*Figure S6: Lifetime mapping on a Nano-diamond cluster. a) TEM images of the cluster, the white square represents the scanned area. Insert: Scanning probe over the region of interest taken with the TEM CCD camera. b) Intensity map extracted from the decay trace map by summing all the histogram bins over a 500 ns window, pixel size 10 nm. c) Lifetime map after binning four adjacent pixels (effective pixel size 20 nm). At each pixel a fit of the decay trace has been done and the extracted lifetime reported on the map. Pixels where the signal was too low to extract a lifetime are masked. d) Error of the lifetime from the square root of the covariance of the fit.*